\documentclass[aps,prd,preprint,groupedaddress,nofootinbib]{revtex4-2}

\usepackage[dvipdfmx]{graphicx,hyperref}
\usepackage{enumitem}
\usepackage{amsmath,amssymb,amsthm,mathtools}
\usepackage{cases}
\usepackage{comment}
\usepackage{color}
\usepackage{physics}
\usepackage{ulem}

\newcommand{\ie}{\textit{i}.\textit{e}. }
\newcommand{\secref}[1]{\S\ref{#1}}

\begin{document}

	\preprint{KUNS-2926}

	\title{Causality bounds chaos in geodesic motions}

	\author{Koji Hashimoto}
		\email{koji@scphys.kyoto-u.ac.jp}
	\author{Kakeru Sugiura}
		\email{sugiura@gauge.scphys.kyoto-u.ac.jp}
	\affiliation{Department of Physics, Kyoto University, Kyoto 606-8502, Japan
		\vspace*{40mm}
	}

	\vspace*{20mm}

	\begin{abstract}
		Predictability is ensured by causality while lost in chaos. 
		To reconcile these two popular notions, we study chaos in geodesic motions in generic curved spacetimes with external potentials, where causality is controlled by a scalar potential. 
		We develop a reparametrization-independent method to analytically estimate the Lyapunov exponent $\lambda$ of a particle motion.
		We show that causality gives the universal upper bound $\lambda\propto E\ (E\rightarrow\infty)$, which coincides with the chaos energy bound proposed by Murata, Tanahashi, Watanabe, and one of the authors (K.H.).
		We also  find that the chaos bound discovered by Maldacena, Shenker, and Stanford	can be violated in particular potentials, even with	causality. 
		Our estimates, although waiting for numerical confirmation, reveal the hidden nature of physical theories: causality bounds chaos.
	\end{abstract}

	
	\maketitle
	
	\newpage
	
	\tableofcontents
	
	\section{\label{sec:intro}Introduction}
		Causality and chaos --- these two popular notions in physics are antithetical, or even conflicting, to each other. Causality is the foundation for predictability in physics, while chaos measures how the predictability of dynamical systems is lost in time evolution. How do these notions reconcile with each other, like Bohr's complementarity in quantum mechanics?
	
		Causality is better defined in relativistic theories, and plays a key role in general relativity. 
		The peculiarity is enlightened when curved spacetimes have event horizons, in the initial value problems using Cauchy surfaces. 
		Causality principle even prohibits some spacetimes, for example, the G\"{o}del solution \cite{RevModPhys.21.447} in the general relativity is regarded as unphysical because of the presence of closed timelike curves, which lead to pathological occurrences such as the grandfather paradox. 
		Hence causality principle has been used in confirming the relativistic consistencies, for instance, of modified gravity theories \cite{Camanho:2014apa,Adams:2006sv,Shore:2007um,Hollowood:2007kt,Edelstein:2016nml}.

		Chaos, remember, was born in Poincar\'e's analysis of three-body problems, regarding gravitational motion of stars in spacetimes. Consequently, the chaos of the geodesic motions in curved spacetime has been investigated for long decades and applied to the explications of the properties of spacetimes \cite{Gibbons:2008hb,Barbon:2011pn,Barbon:2011nj,Barbon:2012zv}. 
		For example, the Lyapunov exponent of null rays was found to relate to quasi-normal mode frequencies of black holes \cite{Cardoso:2008bp}. 
		Chaos of particle motions around black holes has been studied extensively \cite{Sota:1995ms,Suzuki:1996gm}, and it was revealed that the Lyapunov exponent $\lambda$ for the motions near the horizon becomes $\lambda=\kappa$ \cite{Hashimoto:2016dfz}, where $\kappa$ is the surface gravity of the black hole. 
		
		Therefore, for the investigation of possible relations between causality and chaos, geodesic motions of a particle in generic curved spacetimes are a suitable ground. 
		In classical relativistic systems, causality bounds particle motions in/on light cones, which should restrict the complexity of motions. 
		This leads us to the expectation that causality generically gives some fundamental bound on the complicated time evolution of degrees of freedom, chaos. 
		
		In this respect, we prepare two proposed upper bounds on the Lyapunov exponents of chaos. 
		The first one is the renowned chaos bound proposed by Maldacena, Shenker, and Stanford (MSS) \cite{Maldacena:2015waa}. It asserts that the Lyapunov exponent of out-of-time-ordered correlators (OTOCs) \cite{Larkin,Shenker:2013pqa} defined in large $N$ quantum field theory is bounded as 
		\begin{align}
			\lambda\leq\frac{2\pi T}{\hbar},
				\label{MSSbound}
		\end{align}
		where $T$ is the temperature of the system. 
		Originally, this bound was discovered with AdS/CFT correspondence \cite{Maldacena:1997re} through thought experiments of shock waves near black hole horizons \cite{Shenker:2013pqa,Shenker:2013yza}, whose surface gravity is expressed as $\kappa=2\pi T/\hbar$. As the horizon is defined by causality, the MSS chaos bound could be a manifestation of the causality in the bulk curved spacetime, once the holographic principle is assumed.
		
		Another proposal of a universal bound on the Lyapunov exponent is the chaos energy bound \cite{Hashimoto:2021afd}. This conjectures an upper bound of the energy dependence of the Lyapunov exponent $\lambda$ 
		\begin{align}
			\lambda(E)\overset{E\rightarrow\infty}{\propto}E^c,\qquad c\leq1
				\label{energybound}
		\end{align}
		using the energy $E$ of the system, which is consistent in the large $N$ limit with the MSS bound. 
		This bound was motivated by the thermodynamic consistency of OTOCs and is applicable to any classical/quantum Hamiltonian mechanics and field theories, on some reasonable assumptions such as locality. 
		It remains unclear whether this energy bound can be derived from more fundamental principles of physics such as causality.
		
		The purpose of this paper is to find any relation between chaos in relativistic systems and causality, whose significance has tended to be overlooked since causality is assumed for granted. 
		We estimate the Lyapunov exponent of the geodesic motions and elucidate that causality gives the upper bound of its energy dependence, which coincides with the proposed chaos energy bound \eqref{energybound}. 
		
		We consider generic situations in which a particle is moving in a general stationary spherically symmetric spacetime and is coupled to general vector/scalar potentials. There the causality of the particle motion is controlled by the scalar potential $V$ with $p^2=-(m^2+V)$, where $p$ and $m$ are the four-momentum and the mass of the particle, respectively. Requiring that the physical particle should never go back to the past imposes the restriction $m^2+V\geq0$ on allowed scalar potentials. 
		We find that, owing to this restriction stemmed by causality, \textit{the energy bound \eqref{energybound} is always satisfied.}
		On the other hand, we also show that \textit{the MSS bound can be violated under particular potentials, even with the causality restriction.} 
		With these results, we claim that the causality actually provides a bound on chaos, taking the form of the chaos energy bound \eqref{energybound}, while the MSS bound \eqref{MSSbound} is a stronger one possibly dictated by holography.
		
		Note that our analytic study merely estimates, not calculates exactly, the upper bound of the Lyapunov exponent. 
		It points out the possibility of the violation of the MSS bound, however, whether the bound is violated in actual chaotic motions is another issue which we do not address, because the analytic calculation of Lyapunov exponents in our generic situations is quite hard. More numerical works are in need, for verifying the satisfaction/violation of the bounds in each concrete model. 
		In particular, the violation of the MSS bound suggests that, beyond causality, some fundamental principle that rules out those potentials which violate the bound is missing for holographic descriptions of gravity. 
		
		This paper is organized as follows: 
		First, in \secref{sec:method}, we provide a method to analytically estimate the maximal Lyapunov exponent $\lambda$ of geodesic motions in generic spacetimes with vector/scalar potentials. The method is applicable in any parametrization of the geodesic with einbein $e$ and is used to study two chaos bounds, the energy bound and the MSS bound. 
		In \secref{sec:e=1}, we evaluate the Lyapunov exponent in the $e=1$ gauge and investigate the energy bound \eqref{energybound}. We show this bound can be violated by the scalar potential, however, causality restricts its asymptotic behavior in the $E\rightarrow\infty$ limit and protects the bound. 
		In \secref{sec:static}, we convert the Lyapunov exponents in the $e=1$ gauge to those in the static gauge, using the method developed in \secref{sec:method}, and investigate the MSS bound \eqref{MSSbound}. There we show that for particular potentials the Lyapunov exponent $\lambda$ can diverge in the $E\rightarrow\infty$ limit and trivially violate the MSS bound. 
		The final section is for our conclusion and discussion.

	\section{\label{sec:method}Reparametrization-independent estimation of the Lyapunov exponent}
		In this section, we provide a method to analytically estimate the maximal Lyapunov exponent $\lambda$ of geodesic motions in generic spacetimes. 
		First, in \secref{subsec:sbh_formulation}, we illustrate how to estimate $\lambda$ in the simplest case of the 4-dimensional Schwarzschild spacetime. 
		Then in \secref{subsec:general_formulation} we describe our method for generic situations in which a particle is moving in a general stationary spherically symmetric spacetime and is coupled to general vector/scalar potentials. The method is applicable in any parametrization of the geodesic and will be used to study the chaos bounds in the later sections.

		\subsection{\label{subsec:sbh_formulation}Example: Schwarzschild spacetime}
			In generic situations the Lyapunov exponent in chaotic motions is measured numerically, while our interest in the chaos bounds needs analytic expression for the exponent. 
			Hence in this paper we rely on an intuitive estimation method for the Lyapunov exponent focusing on the accelerated particle motion at separatrices in the phase space. This amounts to the estimation of a curvature of the effective potential at its local maxima.
			The method was used in \cite{Hashimoto:2016dfz} and subsequent papers in the context of chaotic motions in spacetimes and the AdS/CFT correspondence.

			In this subsection, as an introductory example, let us consider the radial motion of a free particle in the Schwarzschild spacetime with four dimensions. 
			The illustration is for two purposes: a demonstration in the simplest and well-known example, and a caution for an unreasoning use of popular conservation laws without looking at the worldline gauge. 
			In fact, we will find in \secref{subsec:general_formulation} that the calculation in this subsection depends on a special choice of the parametrization of the geodesic.

			The metric in the 4-dimensional Schwarzschild spacetime is given by
			\begin{align}
				ds^2=-\left(1-\frac{2M}{r}\right)dt^2+\frac{dr^2}{1-2M/r}+r^2d\varphi^2,
			\end{align}
			where we put $\theta=\pi/2$ from the spherical symmetry. We have conserved quantities 
			\begin{align}
				E&=\left(1-\frac{2M}{r}\right)\,p^t,\\
				L&=r^2p^\varphi,
			\end{align}
			where they are the Killing energy and the Killing angular momentum, respectively, and $p^\mu$ is the four-momentum of a particle. 
			The standard normalization of four-momentum
			 \begin{align}
				-m^2 &=p^2
						\label{pconsev}
			\end{align}
			leads to
			\begin{align}
				-m^2 
					&=-\frac{E^2}{1-2M/r}+\frac{\dot{r}^2}{1-2M/r}+\frac{L^2}{r^2},
			\end{align}
			which is rephrased as
			\begin{align}
				E^2=\dot{r}^2+\left(m^2+\frac{L^2}{r^2}\right)\left(1-\frac{2M}{r}\right).
					\label{sbh_consev}
			\end{align}
			We define the effective potential as 
			\begin{align}
				V_{\text{eff}}(r)&:=\left(m^2+\frac{L^2}{r^2}\right)\left(1-\frac{2M}{r}\right),
					\label{sbh_Veff} 
			\end{align}
			then the energy integral \eqref{sbh_consev} describes a particle moving in the potential $V_\text{eff}$.
			
			To estimate the Lyapunov exponent, we closely look at the motion around a stationary point $r=r_0$ of the potential $V_\text{eff}(r)$. We expand the radial coordinate as $r=r_0+\varepsilon$,	then the energy integral \eqref{sbh_consev} can be cast into the form
			\begin{align}
				E^2-V_\text{eff}(r_0)=\dot{\varepsilon}^2+\frac{1}{2}V_\text{eff}''(r_0)\varepsilon^2+\order{\varepsilon^3}.
			\end{align}
			When $V_\text{eff}''(r_0)$ is negative, the system is described by an inverted harmonic oscillator, which has solutions of the form $\varepsilon=Ae^{\lambda\tau}$, where $\lambda := \sqrt{-V_\text{eff}''(r_0)/2}$. This $\lambda$ is interpreted as the estimation of the Lyapunov exponent of the geodesic motion, which is given in the case of the Schwarzschild spacetime as
			\begin{align}
				\lambda	&=\sqrt{-\frac{L^2}{r^5}\left(-\frac{2Mm^2}{L^2}r^2+3r-12M\right)}.
					\label{sbh_lyapunov}
			\end{align}

		\subsection{\label{subsec:general_formulation}Generic spacetimes}
			In the analysis in \secref{subsec:sbh_formulation}, we borrowed the popular conserved quantities and the four-momentum \eqref{pconsev}. However, the gauge dependence of the worldline time $\tau$ was hidden in such a formulation. 
			Since the Lyapunov exponent $\lambda$ is defined through the form $r \sim e^{\lambda\tau}$, it should depend on the worldline parametrization of the geodesic. 
			In this subsection, we describe a reparametrization-independent estimation of the Lyapunov exponent applicable for more general situations: we formulate geodesic motions without fixing the gauge of the worldline time $\tau$, and apply the inverted-harmonic-oscillator approximation to estimate the Lyapunov exponent.
			
			Let us consider a relativistic particle with mass $m$ and charge $q$ in a $D$-dimensional stationary spherically symmetric spacetime with the metric given by 
			\begin{align}
				ds^2=-f(r)dt^2+\frac{dr^2}{g(r)}+r^2d\varphi^2,
					\label{metric}
			\end{align}
			where we put the $D-3$ angular coordinates to take the value $\pi/2$ from the spherical symmetry, and name the remainder $\varphi$. 
			Note that we do not specify the gravity theory. 
			
			The Lagrangian of the particle coupled to an electromagnetic potential $A_\mu=A_\mu(r)$ and a scalar potential $V=V(r)$ is given by 
			\begin{align}
				\mathcal{L}=\frac{1}{2e}g_{\mu\nu}\dot{X}^\mu\dot{X}^\nu-\frac{em^2}{2}+qA_\mu\dot{X}^\mu-\frac12 eV,
					\label{lagparticle}
			\end{align}
			where dots denote the derivative with respect to the geodesic parameter $\tau$ and $e=e(\tau)$ is an einbein, which is the worldline metric.
			The equations of motion derived from this Lagrangian are 
			\begin{align}
				\ddot{X}^\mu+\Gamma^\mu_{\ \nu\rho}\dot{X}^\nu\dot{X}^\rho-\frac{\dot{e}}{e}\dot{X}^{\mu}&=eqF^\mu_{\ \nu}\dot{X}^\nu-\frac12 e^2V^{;\mu},
					\label{geodesic}\\
				-g_{\mu\nu}\dot{X}^\mu\dot{X}^\mu&=e^2(m^2+V).
					\label{constraint}
			\end{align}
			Here \eqref{geodesic} is the geodesic equations, and \eqref{constraint} is the one corresponding to	\eqref{pconsev} in \secref{subsec:sbh_formulation}. 
			We can also find conserved quantities 
			\begin{align}
				E&=\frac{f}{e}\dot{t}-qA_t,
					\label{energy}\\
				L&=\frac{r^2}{e}\dot{\varphi}+qA_\varphi,
					\label{angmom}
			\end{align}
			whose conservation follows from the geodesic equations. Note that these are invariant under reparametrization.
			
			Let us assume the solution of the form $r=r_0+Ae^{\lambda\tau}$ and $e=e(r_0)+e'(r_0)Ae^{\lambda\tau}$ expected in the inverted-harmonic-oscillator approximation and determine $r_0$ and $\lambda$ from the equations of motion (\refeq{geodesic}), (\refeq{constraint}). Substituting these solutions give 
			\begin{align} 
				0= 
				\left[ 
				\frac{f'}{2f^2}(E+qA_t)^2-\frac{1}{r^3}(L-qA_\varphi)^2 
				-q\frac{A'_t}{f}(E+qA_t)-q\frac{A'_\varphi}{r^2}(L-qA_\varphi)+\frac12 V'
				\right]_{r=r_0},
					\label{e_A0}
			\end{align}
			\begin{align}	
				\frac{\lambda^2}{e(r_0)^2}=
					&\left[
					-\left(\frac{f''g}{2f^2}-\frac{f'^2g}{f^3}\right)(E+qA_t)^2 
					-\frac{3g}{r^4}(L-qA_\varphi)^2
					-\frac12 gV''
					\right.
					\nonumber \\
					&
					\left.
					+qg\left(\left(\frac{A''_t}{f}-\frac{2A'_tf'}{f^2}\right)(E+qA_t)+\frac{qA'^2_t}{f}+\left(\frac{A''_\varphi}{r^2}-\frac{4A'_\varphi}{r^3}\right)(L-qA_\varphi)-\frac{qA'^2_\varphi}{r^2}\right) \right]_{r=r_0},
					 	\label{e_A1}
			\end{align}
			\begin{align}
				\Big[(L-qA_\varphi)^2\Big]_{r=r_0}\simeq&\left[r^2\left(\frac{(E+qA_t)^2}{f}-\left(m^2+V\right)\right)\right]_{r=r_0}
					\label{nonfree_LE}
			\end{align}
			in the $A\ll r_0$ limit, where we have assumed $\dot{r}\ll m$ in \eqref{nonfree_LE} so that the particle motion is only near the extremum. Here, $r_0$ determined by $\order{A^0}$ of the geodesic equation \eqref{e_A0} is identified as the position of the extrema of the effective potential, and $\lambda$ by $\order{A^1}$ \eqref{e_A1} is the Lyapunov exponent. The energy integral \eqref{nonfree_LE} relates $L$ with $E$ at $r=r_0$, which will be used in later sections.
			As a consistency check, we note that \eqref{e_A0} does not depend on $e(\tau)$, which should be the case because the location of the extrema should be independent of the intrinsic parameter $\tau$.
			
			In this manner, we have obtained the analytic estimation \eqref{e_A1} of the Lyapunov exponent for the geodesic motions in generic spacetimes with the external fields, without fixing the gauge of the worldline parameter.
			As a by-product, we find that the formulation used in \secref{subsec:sbh_formulation} implicitly assumed the $e=1$ gauge, as is seen in the comparison between \eqref{pconsev} and \eqref{constraint}.

			Our result shows that the einbein dependence of $\lambda$ appears, remarkably, only through the form of $\lambda/e$. This characteristic dependence will be used to convert $\lambda$ between different gauge choices in \secref{sec:static}.

	\newpage
	\section{\label{sec:e=1}Causality bounds the energy dependence of chaos}
		In this section, we elucidate that causality gives the chaos energy bound \eqref{energybound} on the energy dependence of the Lyapunov exponent $\lambda$. 
		We choose a natural gauge $e=1$ in which the gauge choice does not refer to the conserved quantities such as energy. 
		Although the expression \eqref{e_A1} appears complicated, we find that the high energy limit can simplify the argument.
		We begin in \secref{subsec:sbh_estimation} with the Schwarzschild case again and show that $\lambda$ is proportional to $E$ in the $E\rightarrow\infty$ limit, namely the chaos energy bound \eqref{energybound} is saturated. 
		Then in \secref{subsec:general_estimation} we investigate the same limit in our generic situations formulated in \secref{subsec:general_formulation}. The important point is that the scalar potential $V$ controls the causality of the particle via \eqref{constraint}. To reconcile the potential with the causality we impose 
		\begin{align*}
			m^2+V(r)\geq0,
		\end{align*}
		and reveal that this restriction gives an upper bound to the asymptotic order of $V$, which results in the chaos energy bound \eqref{energybound}.

		\subsection{\label{subsec:sbh_estimation}Saturation of the chaos energy bound in the Schwarzschild spacetime}
			We have analytically estimated the Lyapunov exponent $\lambda$ \eqref{sbh_lyapunov} for the geodesic motions in the Schwarzschild spacetime in \secref{subsec:sbh_formulation}.
			Here we prove that it saturates	the chaos energy bound \eqref{energybound}. We find that the $E\rightarrow\infty$ limit resolves the complicated energy dependence of $\lambda$ and extracts its universal behavior.
			
			The extremum of the effective potential at which $\lambda$ is estimated is determined by the zero of the potential slope
			\begin{align}
				V_\text{eff}'(r)&=-\frac{2L^2}{r^4}\left(-\frac{Mm^2}{L^2}r^2+r-3M\right).
					\label{sbh_Veff'}
			\end{align}
			Let us start with the massless particle case for which this equation is simplified. In fact, $V_\text{eff}$ has the local maximum at $r_0=3M$. The relation \eqref{sbh_consev} reduces to $E^2\simeq L^2/27M^2$ at $r=r_0$ when we assume $\dot{r}\ll m$, then we find
			\begin{align}
				\lambda^2=\frac{E^2}{3M^2}.
					\label{sbh_massless}
			\end{align}
			In particular, this result shows $\lambda\propto E^1$, and the chaos energy bound \eqref{energybound} is saturated. 
			We note that we have not taken the $E\rightarrow\infty$ limit.	
			
			For the massive particle, $V_\text{eff}$ has the local maximum at the inner one of two stationary points
			\begin{align}
				r_0=\frac{L^2}{2Mm^2}\left(1-\sqrt{1-\frac{12M^2m^2}{L^2}}\right).
			\end{align}
			We note	$r_0>2M$ (in the case of $\abs{L}\geq2\sqrt{3}M$ in which $V_\text{eff}'(r)=0$ has real solutions). 
			Then we find
			\begin{align}
				E^2&\simeq\left(m^2+\frac{L^2}{r^2}\right)\left(1-\frac{2M}{r}\right) \nonumber\\
				&=m^2\left(1-\frac{(1+u)(1-2u)}{9(1-u)}\right)
			\end{align}
			at $r=r_0$, where $u:=\sqrt{1-12M^2m^2/L^2}$. 
			Solving this about $u$ offers
			\begin{align}
				u&=\frac{9(E/m)^2-8}{4}\left(\sqrt{1+\frac{8}{9(E/m)^2-8}}-1\right) \nonumber\\
				&=1-\frac{2}{9}\left(\frac{m}{E}\right)^2-\frac{16}{27}\left(\frac{m}{E}\right)^4+\mathcal{O}\left(\frac{m}{E}\right)^6
			\end{align}
			in the expansion for $E^2\gg m^2$.
			Now we have $r_0=6M/(1+u)$, thus $\lambda$ is calculated as
			\begin{align}
				\lambda^2&=\frac{m^2}{108M^2}\frac{u(1+u)^3}{1-u} \nonumber\\
				&=m^2\left(\frac{1}{3M^2}\left(\frac{E}{m}\right)^2-\frac{29}{27M^2}+\mathcal{O}\left(\frac{m}{E}\right)^2\right) \nonumber\\
				&\overset{E\rightarrow\infty}{\longrightarrow}\frac{E^2}{3M^2}.
					\label{sbh_massive}
			\end{align}
			This aymptotic value coincides with the massless case \eqref{sbh_massless}, and the energy bound is saturated again.
			
			Here we have shown that the chaos energy bound is saturated irrespective of the particle mass, even though the $L$ dependence of $r_0$ for the massive particle makes the $E$ dependence of $\lambda$ complicated\footnote{We can prove that, in any stationary spherical symmetric spacetime, $r_0$ for the free massless particle cannot depend on $L$ (so $\lambda$ is proportional to $E$), while for the massive one it can.}. 
			This simplification suggests that the $E\rightarrow\infty$ limit unveils the existence of the universal energy bound of chaos \eqref{energybound}.

		\subsection{\label{subsec:general_estimation}Causality ensures the chaos energy bound in generic spacetimes}
			We investigate the $E\rightarrow\infty$ limit of $\lambda$ for generic situations \eqref{e_A1} in the $e(\tau)\equiv1$ gauge. This gauge choice is natural for two reasons: first, it is equivalent to taking the geodesic parameter as $\tau=(\text{proper time})/m$ (or its appropriate massless limit), and second, it does not refer to the integral constants (such as energy) which do not appear in the Lagrangian.	
			We show that the scalar potential can generically violate the chaos energy bound \eqref{energybound}, however, causality restricts its asymptotic behavior in the $E\rightarrow\infty$ limit and protects the universal bound. We also find the bound is always saturated in the potentials which vanish asymptotically. 
			Another gauge choice will be studied in the next section.
			
			In general, $r_0$ determined by \eqref{e_A0} must depend on $E$ complicatedly. Thus, as in \eqref{sbh_massive}, we adopt the $E\rightarrow\infty$ limit to extract the universal asymptotic behavior of the Lyapunov exponent $\lambda$ \eqref{e_A1}. 
			We analyze the high energy behavior by denoting the energy exponents of the quantities showing up in the estimation of the Lyapunov exponent. 
			For $r_0$, we assume
			\begin{align}
				r_0\overset{E\rightarrow\infty}{\longrightarrow}\order{E^n}\qquad(n\in\mathbb{R})
			\end{align}
			for simplicity. 
			At high energy, while for $n=0$ the functions of $r$ such as $f, g, A$, and $V$ are $\order{E^0}$, for the case of $n\neq0$ they can have nontrivial $E$ dependence depending on their asymptotic form. We need to determine the form of these functions. 
			
			First, from now on, let us assume
			\begin{align}
				f(r),g(r)\overset{r\rightarrow\infty}{\longrightarrow}1-\frac{2\Lambda}{(D-1)(D-2)}r^2+\order{\frac{1}{r}}
			\end{align}
			where $\Lambda$ is the cosmological constant. 
			The right hand side means that the geometry asymptotes to maximally symmetric spacetimes (Minkowski, de Sitter or anti-de Sitter). 
			In addition, we consider only $n\geq0$, since otherwise in the high energy limit the orbit may fall into the event horizon for the case of black hole spacetimes.
			Next, let us consider vector potential $A_\mu$. If we assume
			\begin{align}
				A_\mu&\overset{r\rightarrow\infty}{\longrightarrow}\order{r^k}\qquad(k\in\mathbb{R})
					\label{A_constraint}\\
				&\overset{E\rightarrow\infty}{\longrightarrow}\order{E^{kn}},
			\end{align}
			then we get 
			\begin{align}
				kn\leq1;
			\end{align}
			otherwise, $E$ is negligible compared to $A_t$, so \eqref{e_A0} which determines $r_0$ becomes independent of $E$ and contradicts $n>0$. Hence we find $E+qA_t\rightarrow\order{E^1}$.
			Finally for the scalar potential $V$, which is the most important for our causality analysis, we assume 
			\begin{align}
				V(r)&\overset{r\rightarrow\infty}{\longrightarrow}r^l \nonumber \\
				&\overset{E\rightarrow\infty}{\longrightarrow}\order{E^{ln}}.
					\label{Vasympt}
			\end{align}
			Then we find that \eqref{e_A0} puts no restriction on $l$, however, the causality with \eqref{constraint} does bound $l$: demanding physical particles never to be spacelike, that is, $-g_{\mu\nu}\dot{X}^\mu\dot{X}^\mu\geq0$ for $\forall r$ along with the geodesic, gives the restriction 
			\begin{align}
				m^2+V(r)\geq0
					\label{causality}
			\end{align}
			on $V$. Then, since $(L-qA_\varphi)^2$ in \eqref{nonfree_LE} is positive, $ln$ is bounded by the asymptotic energy exponent of $(E+qA_t)^2/f$. This bound depends on the behavior of $f(r)$ in the $E\rightarrow\infty$ limit or the asymptotic geometry of spacetimes, thus let us look at it carefully in order.
			
			For the asymptotically flat case, we have $\Lambda=0$ and $f,g\rightarrow1$. With 
			$n>0$, we have $E^2/f\rightarrow\order{E^2}$ and \eqref{nonfree_LE} with causality \eqref{causality} bounds $V$ as 
			\begin{align}
				ln\leq2.
				\label{constraint_n_flat}
			\end{align}
			Then the energy dependence of the terms in the right hand side of \eqref{e_A0} is found as follows: 
			\begin{align}
				(\text{first term})&=\order{E^{2(1-n)}},\\
				(\text{second term})&=\order{E^{2-n}},\\
				(\text{third term})&=\order{E^{1+kn-n}}\leq\order{E^{2-n}},\\
				(\text{fourth term})&=\order{E^{1+kn-2n}}\leq\order{E^{2-2n}},\\
				(\text{fifth term})&=\order{E^{(l-1)n}}\leq\order{E^{2-n}}.
			\end{align}
			If $kn<1$ and $ln<2$, we find $n=0$; otherwise, in the $E\rightarrow\infty$ limit, only the second term is dominant in \eqref{e_A0} so $r_0$ becomes independent of $E$ and contradicts $n>0$. In the $n=0$ case, only the first and second terms are dominant in \eqref{e_A1} and $\lambda\rightarrow\order{E^1}$ is shown. Meanwhile, if $kn=1$ ($ln=2$), the third (fifth) term in \eqref{e_A0} is as the same order as the second, and $r_0$ can depend on $E$ nontrivially. Then we find $\lambda\rightarrow\order{E^{1-n}}$ and the energy bound is satisfied (but not saturated in any cases).
			
			For the asymptotically (anti-)de Sitter case, we have $\Lambda\neq0$ and $f,g\rightarrow\order{E^{2n}}$. With 
			$n>0$, we have $E^2/f\rightarrow\order{E^{2(1-n)}}$ and \eqref{nonfree_LE} with causality \eqref{causality} bounds $V$ as 
			\begin{align}
				(l+2)n\leq2,\qquad n\leq1.
					\label{constraint_n_ads}
			\end{align}
			Then the energy dependence of the terms in \eqref{e_A0} is as follows: 
			\begin{align}
				(\text{first term})&=\order{E^{2-n}},\\
				(\text{second term})&=\order{E^{2-n}},\\
				(\text{third term})&=\order{E^{1+kn-n}}\leq\order{E^{2-n}},\\
				(\text{fourth term})&=\order{E^{1+kn-n}}\leq\order{E^{2-n}},\\
				(\text{fifth term})&=\order{E^{(l+1)n}}\leq\order{E^{2-n}}.
			\end{align}
			If $kn<1$ and $(l+2)n<2$, we find $n=0$; otherwise, in the $E\rightarrow\infty$ limit, only the first and second term is dominant in \eqref{e_A0} so $r_0$  becomes independent of $E$ since $(E+qA_t)^2$ is eliminated as an overall factor and contradicts $n>0$. Then $\lambda\rightarrow\order{E^1}$ is shown as the flat case. Meanwhile, if $kn=1$ ($(l+2)n=2$), the third and fourth (only fifth) term are as the same order as the first and second, and $r_0$ can depend on $E$ nontrivially. Then we find $\lambda\rightarrow\order{E^{1-n}}$ and the energy bound is satisfied in any cases again.
			
			We have evaluated the energy dependence of the Lyapunov exponent $\lambda$ in the $E\rightarrow\infty$ limit in both asymptotically flat/(A)dS cases,
			and we got
			\begin{widetext}
				\begin{align}
					\lambda_{e=1}&\longrightarrow
					\begin{cases}
						\order{E^1}\qquad(kn<1\ \text{and}\ ln<2,\ \text{hence}\ n=0)\\
						\order{E^{1-n}}\qquad(kn=1\ \text{or}\ ln=2,\ n>0)
					\end{cases}
					\qquad(\Lambda=0),
						\label{lambda_e=1_flat}\\
					\lambda_{e=1}&\longrightarrow
					\begin{cases}
						\order{E^1}\qquad(kn<1\ \text{and}\ (l+2)n<2,\ \text{hence}\ n=0)\\
						\order{E^{1-n}}\qquad(kn=1\ \text{or}\ (l+2)n=2,\ 0<n\leq1)
					\end{cases}
					\qquad(\Lambda\neq0).
						\label{lambda_e=1_ads}
				\end{align}
			\end{widetext} 
			Without the causality constraint, the asymptotic energy exponent $l$ of $V(r)$ can be arbitrarily large, and the chaos energy bound is violated. 
			On the other hand, we have discovered that causality gives the upper bound on $l$ and protects the chaos energy bound\footnote{In particular, popular causal setups of the gravity-matter actions may provide potentials which vanish in $r\rightarrow\infty$ \ie $k,l<0$. 
			Our analysis given in this section includes those cases, and leads to the saturation of the chaos energy bound.}.
			Since causality is one of the most fundamental principles of physics, our result suggests that causality ensures the universal bound \eqref{energybound} for any chaotic system.

	\section{\label{sec:static}Causality alone cannot avoid violation of the MSS bound}
		We have evaluated above the Lyapunov exponent $\lambda$ in the $e=1$ gauge and revealed that the causality of the particle ensures the universal bound \eqref{energybound} on the energy dependence of chaos. 
		In this section, in contrast, we study another chaos bound: the MSS bound \eqref{MSSbound}.
		Since this bound is motivated by the AdS/CFT correspondence, the Lyapunov exponent needs to be measured in the static gauge. We make a gauge transformation upon the result given in the previous section, and show that at high energy the MSS bound is violated.
		
		We first remind the readers that the MSS bound is for the boundary theory of the AdS/CFT correspondence, thus the Lyapunov exponent $\lambda$ in the expression of the MSS bound needs to be measured with the boundary time $t$. The boundary time corresponds to the target space time on the gravity side. Thus we need to convert $\lambda$ in the $e=1$ gauge to that in the static gauge using the reparametrization-independent representation \eqref{e_A1}. 
		
		Under this reparametrization, $r_0$ is invariant and the conditions on $n, k$, and $l$ are the same as those in \secref{subsec:general_estimation}. The transformation law is simply 
		\begin{align}
			\lambda_\text{static}=\left.e(\tau=t(\tau))\right|_{r_0}\lambda_{e=1}.
		\end{align}
		Because, in the static gauge, we obtain
		\begin{align}
			e(\tau=t(\tau))&=\frac{f(r)}{E+qA_t}
		\end{align}
		from \eqref{energy}, we find that $\left.e(\tau=t)\right|_{r_0}$ is $\order{E^{-1}}$ in the asymptotically flat case and $\order{E^{2n-1}}$ in the asymptotically (A)dS case. Hence the energy dependence of $\lambda$ becomes as follows: 
		\begin{widetext}
		\begin{align}
			\lambda_\text{static}&\longrightarrow
			\begin{cases}
				\order{E^0}\qquad(kn<1\ \text{and}\ ln<2,\ \text{hence}\ n=0)\\
				\order{E^{-n}}\qquad(kn=1\ \text{or}\ ln=2,\ n>0)
			\end{cases}
			\qquad(\Lambda=0),
			\label{lambda_static_flat}\\
			\lambda_\text{static}&\longrightarrow
			\begin{cases}
				\order{E^0}\qquad(kn<1\ \text{and}\ (l+2)n<2,\ \text{hence}\ n=0)\\
				\order{E^{n}}\qquad(kn=1\ \text{or}\ (l+2)n=2,\ 0<n\leq1)
			\end{cases}
			\qquad(\Lambda\neq0).
			\label{lambda_static_ads}
		\end{align}
		\end{widetext}
		Among the various cases, we find that in the last case, for $\Lambda\neq0$ with $kn=1$ or $(l+2)n=2$, the exponent $\lambda$ diverges in the $E\rightarrow\infty$ limit and violates the MSS bound: since the upper bound $2\pi T/\hbar$ in the MSS bound is described only by the quantities of spacetimes, this divergence obviously violates the bound. 
		
		The violation condition we have found here leads us to state that spacetimes with external potentials having such values of $k$ or $l$, which are still causal, are not realized holographically\footnote{It is amusing to note that this divergence of $\lambda$ can occur particularly in the asymptotically AdS ($\Lambda<0$) case, where the MSS bound was motivated through the AdS/CFT correspondence.}. 
		Causality is not strong enough to make sure the holographic principle.

	\section{\label{sec:conclusion}Conclusion and discussion}
		We have considered the geodesic motions in generic situations in which a particle is moving in a general stationary spherically symmetric spacetime with the metric \eqref{metric} and is coupled to general vector/scalar potentials \eqref{lagparticle}, and have investigated the bound on the Lyapunov exponent $\lambda$ imposed by causality. 
		In \secref{sec:method}, we estimated $\lambda$ as \eqref{e_A1} without fixing the gauge of the worldline time. 
		In \secref{sec:e=1}, we described our main results that causality of the particle motion gives a universal upper bound on the energy dependence of $\lambda$, which coincides with the proposed chaos energy bound \eqref{energybound}. It was important there that causality was reflected in \eqref{constraint} and gave the restriction \eqref{constraint_n_flat}, \eqref{constraint_n_ads} on the asymptotic form of the scalar potential. 
		In contrast, in \secref{sec:static}, we pointed out the celebrated MSS bound could be violated in the particular potentials \eqref{lambda_static_ads}, even with such a causal restriction. 
		This suggests that, beyond causality, some fundamental principle that rules out those potentials which violate the bound is needed for holographic descriptions of gravity.

		Let us make a brief comment on the gauges we chose in this paper. For that purpose, we remind the readers of 
		 another perspective in the study of chaos in gravity theories: chaos in the time evolution of the gravitational field itself, which plays an important role in cosmology \cite{BARROW19821,hobill1994deterministic}. There, subtleties in defining chaos, which is characterized by a positive Lyapunov exponent, are led by the non-invariance of the exponent of the gravitational field under general coordinate transformations \cite{francisco1988qualitative,burd1990numerical,berger1991comments,Hobill_1991,burd1993invariant,Wu:2003pe,Motter:2003jm}. 
		One may notice that this subtlety applies to our study too, because of the non-invariance under the reparametrizations of the worldline time $\tau$. In fact, the energy dependence of the Lyapunov exponent in the $e=1$ gauge \eqref{lambda_e=1_flat}, \eqref{lambda_e=1_ads} has changed as \eqref{lambda_static_flat}, \eqref{lambda_static_ads} in the static gauge. 
		However, we argue for several reasons that, this issue is resolved in our case and the most important energy dependence is the one in the $e=1$ gauge: first, the $e=1$ gauge is equivalent to choosing the gauge of the worldline time for a massive particle as $\tau=\text{(proper time)}/m$, which is the intrinsic time observed in the comoving frame of the particle. Second, an einbein $e$ is unnatural if it depends on the energy $E$, whose conservation is followed from the equations of motion, since the gauge of worldline time must be fixed before integrating the equations. 
		Then the energy dependence of $\lambda$ should not change from \eqref{lambda_e=1_flat}, \eqref{lambda_e=1_ads} and should satisfy the chaos energy bound \eqref{energybound} in the natural gauges.

		Several discussions are in order. Below we make comments on need of (i) numerical simulations, (ii) analysis in less-symmetric cases, and (iii) study in relation to AdS/CFT correspondence. First, 
		our analytic estimate uses the separatrices of the effective potential which the moving particle feels. 
		Although the curvature of the potential produces an exponential behavior of the particle motion, it is merely an estimate of the Lyapunov exponent of the whole motion of the particle. To determine the actual Lyapunov exponent, one needs numerical study for each chosen setup of curved spacetimes and potentials. 
		Our study focuses on universal properties and thus used the analytic estimate, and it is obvious that more numerical study is necessary to check the violation of the MSS bound.
		We emphasize that we have not specified the gravity theory and our results apply to generic spacetimes not only in the general relativity but in its extensions like various modified gravity theories, and numerical/analytic calculation of the Lyapunov exponents in such theories (see \cite{Zhao:2018wkl,Lei:2020clg,Bera:2021lgw,Giataganas:2021ghs,Raffaelli:2020cyl,Gwak:2022xje,Gao:2022ybw,Chen:2022tbb} for relevant study) may lead to further progress in chaos study.
		
		Second, we should add a caveat about the symmetry of our systems and chaoticity. We have focused on geodesic motions in general stationary spherically symmetric spacetimes. This reduces the motion to just a radial one, that is, one-dimensional, thus according to the Poincar\'e-Bendixson theorem there should be no chaos. What we assume is that the separatrices we used would serve as the nest of chaos in generic less-symmetric cases in which chaos is present. The reason why we were forced to use the symmetric spacetimes is that for the analytic evaluation of the Lyapunov exponent we needed the energy exponents of the conserved quantities coming from the spacetime symmetry. In general, chaos emerges in non-integrable systems which have not enough number of conserved quantities, thus it is important to extend our reparametrization-independent estimation of the Lyapunov exponent \eqref{e_A1} to less symmetric spacetimes. 
				
		Third, what is ``some fundamental principle for holographic description" suggested by the violation of the MSS bound? The most conservative candidate is the causality of the boundary theory, not of the bulk gravity theory which we studied. In AdS/CFT correspondence, although a particle in the boundary is described by a local operator, one falling into the bulk is by a composite and non-local operator. Hence the causality constraint of the boundary theory may be nontrivial and more constraining than the bulk one.\footnote{The gravity side could be a string theory in which case the chaos in the bulk is also nonlocal, see \cite{PandoZayas:2010xpn} for the early study of string chaos in the AdS/CFT correspondence.} There is the possibility that this hidden constraint rules out the potentials which violate the MSS bound. It should be important to verify this conjecture.
		
		Finally, let us make a remark on a general perspective on the antithetical nature of causality and chaos. The chaos energy bound \eqref{energybound} was
		motivated by the thermodynamic consistency of the definition of the Lyapunov exponent. Is there anything to do with the relativistic causality which was important for deriving \eqref{energybound} in this paper? This question reminds us of S.~Hawking \cite{Hawking:1975vcx} having discovered that causality division surfaces, {\it i.e.} event horizons, are necessarily thermal. In other words, consistency of thermal ensembles in quantum mechanics is required by the singular causal structure. In this respect, what we see in this paper may be just a tiny part of the physical consistency conditions in the whole theory space unseen to us. It is truly amusing to explore more on yet-to-be-discovered relations between the antithetical notions in physics: causality and chaos.

	\begin{acknowledgments}
		We would like to thank N.~Tanahashi and R.~Watanabe for fruitful discussions, and S.~Iso and K.~Murata for valuable comments. 
		The work of K.~H.~was supported in part by JSPS KAKENHI Grant Numbers 17H06462, 22H01217.
	\end{acknowledgments}

	\bibliography{2205.13818_v2}

\end{document}